\documentclass[10pt,conference]{IEEEtran}
\IEEEoverridecommandlockouts
\usepackage{cite}
\usepackage{amsmath,amssymb,amsfonts,bm}
\usepackage{algorithmic}
\usepackage{graphicx}
\usepackage{textcomp}
\usepackage{xcolor}
\usepackage{physics}
\usepackage{mathtools}

\usepackage{tikz}
\usetikzlibrary{quantikz2}
\usepackage{float}
\usepackage{algorithm}
\usepackage{algorithmic}

\ifCLASSOPTIONcompsoc
 \usepackage[caption=false,font=normalsize,labelfont=sf,textfont=sf]{subfig}
\else
 \usepackage[caption=false,font=footnotesize]{subfig}
\fi

\def\BibTeX{{\rm B\kern-.05em{\sc i\kern-.025em b}\kern-.08em
    T\kern-.1667em\lower.7ex\hbox{E}\kern-.125emX}}

\begin{document}

\title{A double selection entanglement distillation-based state estimator\\
% {\footnotesize \textsuperscript{*}Note: Sub-titles are not captured for https://ieeexplore.ieee.org  and
\thanks{This work was supported by the JST Moonshot R\&D program under Grants JPMJMS226C.}
}

\author{\IEEEauthorblockN{
Joshua Carlo A. Casapao\IEEEauthorrefmark{1},
Ananda G. Maity\IEEEauthorrefmark{1}\IEEEauthorrefmark{2},
Naphan Benchasattabuse\IEEEauthorrefmark{3},\\
Michal Hajdu\v{s}ek\IEEEauthorrefmark{3},
Akihito Soeda\IEEEauthorrefmark{4}\IEEEauthorrefmark{5},
Rodney Van Meter\IEEEauthorrefmark{6},
and David Elkouss\IEEEauthorrefmark{1}}

\IEEEauthorblockA{\IEEEauthorrefmark{1}\textit{Networked Quantum Devices Unit, Okinawa Institute of Science and Technology Graduate University},\\
\textit{Onna-son, Okinawa, Japan}}
\IEEEauthorblockA{\IEEEauthorrefmark{2}\textit{S. N. Bose National Centre for Basic Sciences, Block JD, Sector III, Salt Lake, Kolkata 700 106, India}}
\IEEEauthorblockA{\IEEEauthorrefmark{3}\textit{Graduate School of Media and Governance, Keio University Shonan Fujisawa Campus, Kanagawa, Japan}}
\IEEEauthorblockA{\IEEEauthorrefmark{4}\textit{Principles of Informatics Research Division, National Institute of Informatics, 2-1-2 Hitotsubashi, Chiyoda-ku, Tokyo, Japan}}
\IEEEauthorblockA{\IEEEauthorrefmark{5}\textit{Department of Informatics, School of Multidisciplinary Sciences, SOKENDAI}\\
\textit{(The Graduate University for Advanced Studies),
2-1-2 Hitotsubashi, Chiyoda-ku, Tokyo 101-8430, Japan}}
\IEEEauthorblockA{\IEEEauthorrefmark{6}\textit{Faculty of Environment and Information Studies, Keio University Shonan Fujisawa Campus, Kanagawa, Japan}\\
joshuacarlo.casapao@oist.jp}
}

% \author{\IEEEauthorblockN{Joshua Carlo A. Casapao}
% \IEEEauthorblockA{\textit{Networked Quantum Devices Unit} \\
% \textit{OIST}\\
% Okinawa, Japan \\
% \and
% \IEEEauthorblockN{Ananda G. Maity}
% \and
% \IEEEauthorblockN{Naphan Benchasattabuse}
% \and
% \IEEEauthorblockN{Michal Hajdu\v{s}ek}
% \and
% \IEEEauthorblockN{Akihito Soeda}
% \and
% \IEEEauthorblockN{Rodney Van Meter}
% \and
% \IEEEauthorblockN{David Elkouss}

\maketitle

\begin{abstract}
With the advent of practical quantum communication networks drawing closer, there is a growing need for reliable estimation protocols that can efficiently characterize quantum resources with minimum resource overhead requirement. A novel approach to this problem is to integrate an estimator into an existing network task, thereby removing the need for an additional characterization protocol. In this work, we show that the measurement statistics of a double selection distillation protocol alone can be used to efficiently estimate the Bell-diagonal parameters of the undistilled states, as well as the resulting distilled states after additional post-processing. We also demonstrate that this novel estimator outperforms the previously proposed distillation-based estimator in terms of resource complexity.
\end{abstract}


\section{Introduction}\label{sec:introduction}

Recent experimental progress is enabling quantum communication tasks over near-term quantum networks, paving the way toward the future quantum Internet \cite{kimble2008quantum,Wehner2018quantum,VanMeter2022architecture}. Inevitably, it becomes critical to simultaneously develop robust and efficient methods that certify and characterize the resources used in these networks to ensure proper functioning \cite{Eisert2020}. 
There exist different methods with varying assumptions \cite{Eisert2020}; some examples are partial/full tomographic reconstruction \cite{gross2010quantum,qi2017adaptive,guctua2020fast}, direct fidelity estimation \cite{flammia2011direct}, and randomized benchmarking techniques \cite{knill2008randomized,helsen2022general,helsen2023benchmarking}. It is important that these methods compensate for experimental realities and prioritize both resource efficiency and measurement complexity without sacrificing the amount of learnable information. 

Typically, characterization and certification procedures are conducted independently before implementing other network tasks to ensure the desired condition of the resources used. In contrast, we can consider another avenue in which a characterization method is integrated into an existing network task unrelated to characterization. We expect one such scheme to exploit the behavior of the underlying structure of a quantum resource under a network task, and then post-process any learnable information to obtain a partial or even full characterization. For example, in the realm of quantum error correction, estimators in \cite{Wagner2021optimal,Wagner2022pauli} can use syndrome measurements of stabilizer codes to learn Pauli noise parameters without disturbing the encoded information. By integrating characterization within network tasks, potential methods similar to these can indirectly address the problem of resource consumption by effectively reducing the number of separate protocols to be implemented. %can be effectively reduced.

Recently, it was shown in \cite{Maity23noise,casapao2024distimator} that a \textit{disti}llation-based state esti\textit{mator} can be designed (dubbed Disti-Mator) to efficiently characterize the Bell-diagonal elements of the states prepared for probabilistic distillation protocols. In this particular case, the action of these distillation protocols on the Bell-diagonal elements is exploited. It was shown that partial state estimation can be a natural by-product of post-processing the measurement statistics from distillation. Thus, we can avoid an additional benchmarking procedure whenever distillation is a necessary network task. However, a limitation of the original construction in \cite{casapao2024distimator} is its reliance on three distinct distillation protocols to characterize the Bell-diagonal elements. This may be undesirable if a user wants a consistent supply of high-fidelity states for further network tasks or just wants to avoid reconfiguring a distillation setup during implementation, which directly affects timing coordination.

In this work, we build on top of this previously developed distillation-based estimator. We show that a distillation-based estimator can be designed more compactly by only implementing a double selection distillation protocol and using its measurement statistics, thus introducing a \emph{double selection-based} Disti-Mator. We prove that the measurement statistics obtained from this protocol are enough to efficiently characterize the Bell-diagonal elements of the undistilled states, as well as the distilled states after additional post-processing. Our results are relevant in the context of establishing end-to-end entanglement over real testbeds, given that the proposed estimator avoids a major temporal bottleneck of changing measurement bases for the purpose of estimation.

We structure this work as follows. We first revisit the double selection protocol introduced in \cite{fujii2009entanglement} that we use in our investigation. We then describe the strategy for estimating the Bell-diagonal coefficients, including the relevant concentration bounds. We further show that the resulting distilled states can also be characterized after successfully estimating the noisy input states. We also investigate the number of samples required to estimate these coefficients. We then show that the proposed estimator can outperform the previously proposed Disti-Mator in \cite{casapao2024distimator} relative to the sample complexity. Finally, we provide some numerical tests to assess the utility and feasibility of our estimator during implementation.

%--------------------%

\section{Double selection protocol}\label{sec:doubleselection}

\begin{figure}[tb]
    \centering
    \begin{quantikz}
        \inputD{\rho^{(0)}} & \ctrl{1} & \qw       & \qw       \\[0.1cm]
        \inputD{\rho^{(1)}} & \targ{}  & \targ{}   & \meter{Z} \\[0.1cm]
        \inputD{\rho^{(2)}} & \qw      & \ctrl{-1} & \meter{X}
    \end{quantikz}
    \caption{Double selection distillation protocol \cite{fujii2009entanglement} between two users, where one half of the protocol local to one user is shown. The protocol is implemented with three noisy input states and bilateral quantum operations. The protocol itself is a probabilistic protocol (that is, a successful distillation of a noisy input state is not deterministic) that requires two-way classical communication between the users for postselecting $Z$ and $X$ measurement results.}
    \label{fig:double_selection_circuit}
\end{figure}
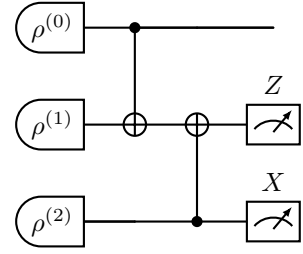

We consider the double selection distillation protocol \cite{fujii2009entanglement} for two parties, Alice and Bob, with half of the protocol illustrated in Figure~\ref{fig:double_selection_circuit}. The protocol is as follows:
\begin{enumerate}
    \item Alice and Bob prepare and share three identical noisy Bell pairs, which we label $\rho^{(0)}$, $\rho^{(1)}$, $\rho^{(2)}$.
    \item Both parties implement a bilateral CNOT on $\rho^{(0)}$ and $\rho^{(1)}$, with the former as the control qubit.
    \item Both parties implement a bilateral CNOT on $\rho^{(1)}$ and $\rho^{(2)}$, with the latter as the control qubit.
    \item They bilaterally measure $\rho^{(1)}$ in the $Z$ basis and $\rho^{(2)}$ in the $X$ basis. Then they classically communicate their measurement results with each other.
    \item Postselection stage: if both confirm coincidences in the $Z$ measurements and in the $X$ measurements, then $\rho^{(0)}$ is successfully distilled and is prepared for the next round. Otherwise, $\rho^{(0)}$ is discarded.
\end{enumerate}
Before each subsequent round, the roles of the $X$ and $Z$ bases on the reference frames of the surviving states are inverted. It has been shown in \cite{fujii2009entanglement} that by using two ancillary noisy Bell pairs for postselection, this protocol can remove both phase and bit flip errors more efficiently than the single selection protocols in \cite{Bennett96mixed-state,Deutsch96quantum} that only use a single ancillary pair. Additionally, double selection provides higher-fidelity states much faster after several rounds and can achieve yields comparable to single selection. 

%--------------------%

\section{Main results}

\subsection{Estimation strategy}\label{subsec:strategy}

We propose that both the undistilled and distilled states in a double selection protocol can be estimated using the coincidence probabilities \textit{over a single distillation round}. Here, we consider that the input states $\rho = \overline{\rho} + \rho_{\text{off-diagonal}}$, where $\overline{\rho}$ is in Bell-diagonal form:
\begin{multline}
    \overline{\rho}(\mathbf{q}) \equiv q_1 \ketbra{\Phi^+}{\Phi^+} + q_2 \ketbra{\Phi^-}{\Phi^-}\\
    + q_3 \ketbra{\Psi^+}{\Psi^+} + q_4 \ketbra{\Psi^-}{\Psi^-}.
\end{multline}
We describe $\mathbf{q}\equiv (q_1,q_2,q_3,q_4)$ as the set of convex coefficients that satisfy the normalization $q_1 + q_2 + q_3 + q_4 = 1$. We further assume that $q_1 > 1/2$, reflecting a situation in which the desired distilled states tend towards the $\ket{\Phi^+}$ state. To estimate $\overline{\rho}(\mathbf{q})$, we need to determine three unknowns, which means we need a system of at least three equations. We propose that these three equations correspond to the three coincidence probabilities that we can extract from the protocol: the $Z$-marginal, the $X$-marginal, and the joint coincidences that signal a successful distillation.

Before we proceed, we also clarify the other assumptions that we have set for the estimator. Firstly, we assume that the input Bell pairs are independent and identical. Hence, we will not consider drifting sources. Secondly, we assume that the implementation of the distillation protocol itself is performed perfectly. That is, the quantum devices used in the protocol are not subject to imperfections due to noise. Only the sources are imperfect, producing noisy Bell pairs. There is also no noise acting on the Bell pairs whenever they are idle in a quantum memory. Although these assumptions limit the scope of this work, we believe that the observations we make here can easily be extended to a more realistic scenario, which we keep open for future studies.

We are now ready to describe the relevant coincidence probabilities, and the rest of our estimation strategy. For our investigation, we consider the transformation
\begin{equation}
    x_i \equiv q_1 + q_{i+1}, \qquad i\in\{1,2,3\},
\end{equation}
so that
\begin{equation}\label{eqn:q-x_transformation}
    \begin{pmatrix}
        q_1 \\ q_2 \\ q_3 \\ q_4
    \end{pmatrix}
    =\frac{1}{2}\begin{pmatrix}
        -1 &  1 &  1 &  1 \\
         1 &  1 & -1 & -1 \\
         1 & -1 &  1 & -1 \\
         1 & -1 & -1 &  1 
    \end{pmatrix}
    \begin{pmatrix}
        1 \\ x_1 \\ x_2 \\ x_3
    \end{pmatrix},
\end{equation}
and $x_i > 1/2$ for any $i\in\{1,2,3\}$. As we shall immediately see, our choice of this transformation rule allows two of the equations considered here to be single-variable functions instead of being multivariate in $\mathbf{q}$. The coincidence probability on $Z$ alone is described by
\begin{equation}\label{doub_coin_prob_Z}
    p^{(Z)}(x_1) = 3x_1 - 6x_1^2 + 4x_1^3,
\end{equation}
where $1/2 < p^{(Z)} \leq 1$. For the coincidence probability on $X$ alone, we have 
\begin{equation}\label{doub_coin_prob_X}
    p^{(X)}(x_2) = 1-2x_2 + 2x_2^2,
\end{equation}
where $1/2 < p^{(X)} \leq 1$. Finally, for the coincidence probability on both $Z$ and $X$, we have the only remaining multivariate function
\begin{multline}\label{doub_coin_prob_XZ}
    p^{(XZ)}(\mathbf{x})=2x_1 -3x_1^2 +2x_1^3 -x_2 +x_2^2 \\
    + x_3-2x_1x_3 -x_3^2 +2x_1x_3^2,
\end{multline}
where $1/4 < p^{(XZ)} \leq 1$. Here, we use the notation $\mathbf{x} \equiv (x_1,x_2,x_3)$. Notice that over the interval $x_1\in(1/2,1]$, we have the derivative $\partial_{x_1} p^{(Z)}(x_1) = 3(2x_1-1)^2 \geq 0$. This is also the case for $p^{(X)}$ over the interval $x_2\in(1/2,1]$, that is, $\partial_{x_2} p^{(X)}(x_2) = 2(2x_2-1) \geq 0$. Hence, both $p^{(Z)}$ and $p^{(X)}$ are (strictly) monotonically increasing over these intervals. 

It is possible to invert these coincidence probabilities to $\mathbf{x}$. From \eqref{doub_coin_prob_Z} we obtain 
\begin{equation}\label{eqn:doub_x1_noiseless}
    x_1({p}^{(Z)}) = \frac{1}{2}\left(1+\left(2p^{(Z)}-1\right)^{1/3}\right),
\end{equation}
while from \eqref{doub_coin_prob_X} we have
\begin{equation}\label{eqn:doub_x2_noiseless}
    x_2({p}^{(X)}) = \frac{1}{2}\left(1+\left(2p^{(X)}-1\right)^{1/2}\right).
\end{equation} 
Finally, we consider $p^{(XZ)}$. Fixing both $x_1$ and $x_2$, we observe that $p^{(XZ)}$ is quadratic in $x_3$. Substituting \eqref{eqn:doub_x1_noiseless} and \eqref{eqn:doub_x2_noiseless} into \eqref{doub_coin_prob_XZ} and assuming $x_3\in(1/2,1]$, we have
\begin{multline}\label{eqn:doub_x3_noiseless}
    x_3({p}^{(Z)},{p}^{(X)},{p}^{(XZ)}) \\
    = \frac{1}{2}\left[1+\left(\frac{4p^{(XZ)}-2p^{(Z)}-2p^{(X)}+1}{(2p^{(Z)}-1)^{1/3}}\right)^{1/2}\right], 
\end{multline}
where we avoid the condition $p^{(Z)}= 1/2$ by our assumption on $q_1$, and that $x_1$ and $x_2$ uniquely determine $p^{(Z)}$ and $p^{(X)}$, respectively. Unlike the previous expressions, $x_3$ is a complicated function of the three coincidence probabilities. 

Using \eqref{eqn:doub_x1_noiseless}-\eqref{eqn:doub_x3_noiseless}, we can determine the corresponding intermediate estimate $\hat{\mathbf{x}}\equiv (\hat{x}_1,\hat{x}_2,\hat{x}_3)$ by substituting the empirical probabilities $\hat{p}^{(Z)}$, $\hat{p}^{(X)}$, and $\hat{p}^{(XZ)}$ (i.e., the measurement statistics of the protocol). We then transform $\hat{\mathbf{x}}$ into the desired estimate $\hat{\mathbf{q}}\equiv(\hat{q}_1,\hat{q}_2,\hat{q}_3,\hat{q}_4)$ of the vector $\mathbf{q}$ using the transformation rule in \eqref{eqn:q-x_transformation}.

\subsection{Concentration bounds}\label{subsec:concentrationbounds}

We now show the estimation guarantee of our proposed estimator. We provide the bulk of our mathematical proof in the Appendices. We evaluate this guarantee with the trace distance $D(\hat{\rho}(\hat{\mathbf{q}}),\overline{\rho}(\mathbf{q}))$ between the true Bell-diagonal elements $\overline{\rho}(\mathbf{q})$ and our estimation $\hat{\rho}(\hat{\mathbf{q}})$ as our figure of merit. We choose the trace distance as a measure of quality due to its operational interpretation \cite{wilde2013quantum} and its usage in a range of certification protocols for applications in quantum networks \cite{Eisert2020}. 

Central to our proof is to determine the concentration inequalities associated with a successful estimation of $\mathbf{x}$. That is, we want to ensure that with high probability
\begin{equation}\label{eqn:x_assumption_diff_eps}
    \abs{\hat{x}_i - x_i}\leq \varepsilon_i    
\end{equation}
for all $i$, where $\varepsilon_i>0$ is some additive error. We rigorously show these inequalities in Appendices~\ref{app:concbound_x1x2} and~\ref{app:concbound_x3}. In the proof, the concentration bounds are guaranteed by the Hoeffding bounds associated with the measurement statistics, where a confirmed or refuted coincidence is an independent Bernoulli random variable. For the Hoeffding bounds, we let $N$ be the number of triplets of input states ${\rho}^{\otimes 3}$ used in the double selection protocol, corresponding to the $N$ Bernoulli random variable pairs. We point out that we use the same measurement data to estimate all the true coincidence probabilities. 

We note two things. First, the bounds for $x_1$ and $x_2$ are the easiest to show, since each is a univariate function of one of the coincidence probabilities (given in \eqref{eqn:doub_x1_noiseless} and \eqref{eqn:doub_x2_noiseless}). On the other hand, proving the probability bound for $x_3$ is more difficult, since $x_3$ is a multivariate function of all coincidence probabilities (given in \eqref{eqn:doub_x3_noiseless}). Secondly, the additive errors $\varepsilon_i$ are determined by errors $\varepsilon^{(j)}$ associated with the empirical coincidence probabilities $\hat{p}^{(j)}$ obtained during implementation, where $j=X,Z,XZ$.

With the estimation guarantee for $\mathbf{x}$ established, we then propagate the errors in $\hat{\mathbf{x}}$ to obtain a concentration inequality associated with the trace distance between the estimation $\hat{\rho}$ and $\overline{\rho}$. The details are shown in Appendix~\ref{app:conc_trace_distance}. Our estimation guarantee is therefore
\begin{equation}\label{eqn:conc_bound_trace_dist_final}
    \Pr(D(\hat{\rho}(\hat{\mathbf{q}}),\overline{\rho}(\mathbf{q}))\geq \varepsilon_T)\equiv \sum_{i=1}^3 \delta_i\leq \delta_T,
\end{equation}
where the $\delta_i$'s are the respective probability bounds associated with the $x_i$'s, and $\varepsilon_T \equiv \sum_{i=1}^3 \varepsilon_i$ is the sum of the additive errors.

We finally investigate the sample complexity of our estimator. Assuming that the errors $\varepsilon^{(j)}$ chosen in the Hoeffding bounds are the same, we expect that the sample complexity must pessimistically be 
\begin{equation}
    N = \mathcal{O}\left(\frac{\log(1/\delta_T)}{\varepsilon_T^2}\right).
\end{equation}
We provide a short discussion on this in Appendix~\ref{app:complexity}.

We summarize our proposed estimation protocol with the following algorithm.
\begin{flushleft}
\noindent \textbf{Algorithm:} Double selection-based Disti-Mator
\end{flushleft}
\begin{algorithmic}[1]
\renewcommand{\algorithmicrequire}{\textbf{Input:}}
\renewcommand{\algorithmicensure}{\textbf{Output:}}
\REQUIRE $N$ triplets $\rho^{\otimes 3}$ of the noisy Bell pair; desired errors $\bm{\varepsilon}^{(\cdot)}\equiv(\varepsilon^{(Z)},\varepsilon^{(X)},\varepsilon^{(XZ)})$ on the coincidence probabilities.
\ENSURE Estimate $\hat{\mathbf{q}}$; trace distance bound $\varepsilon_T$; failure probability $\delta_T$.
\STATE Perform a single round double selection protocol on the $N$ triplets and then collect the measurement results.
\STATE Calculate the empirical coincidence probabilities $\hat{\mathbf
{p}}\equiv(\hat{p}^{(Z)},\hat{p}^{(X)},\hat{p}^{(XZ)})$. 
\IF {$x_i(\hat{\mathbf
{p}}\pm \bm{\varepsilon}^{(\cdot)})$ is not a real number within $(1/2,1]$ for any $i\in[3]$ (see \eqref{eqn:doub_x1_noiseless}-\eqref{eqn:doub_x3_noiseless})} 
    \STATE $\hat{\mathbf{q}}\gets \mathrm{null}$.
    \STATE $\varepsilon_T \gets 1$.
    \STATE $\delta_T \gets 1$. 
    \PRINT Estimation protocol fails. Increase $N$ and/or change $\bm{\varepsilon}^{(\cdot)}$. 
\ELSE
    \STATE $\hat{x}_i \gets x_i(\hat{\mathbf{p}})$.
    \STATE Calculate $\varepsilon_i$ for all $i\in[3]$ (see Appendices~\ref{app:concbound_x1x2} and~\ref{app:concbound_x3}).
    \STATE Transform $\hat{\mathbf{x}}$ into $\hat{\mathbf{q}}$ using \eqref{eqn:q-x_transformation}.
    \STATE $\varepsilon_T \gets \sum_{i=1}^3 \varepsilon_i$.
    \STATE Calculate probability bounds $\delta_i$ using the Hoeffding bounds associated with $\hat{\mathbf{p}}$ (see Appendix~\ref{app:conc_trace_distance}).
    \STATE $\delta_T \gets \sum_{i=1}^3 \delta_i$.
\ENDIF 
\end{algorithmic}

\subsection{Estimating distilled states}\label{subsec:distilledstates}

We note that the distilled states can be estimated for free given that we only need to propagate the errors on the undistilled states over an application of the entire double selection protocol. We use a similar technique to \cite{casapao2024distimator}, in this case treating the double selection protocol as an action of some quantum instrument on the unmeasured noisy Bell pair. Let $\mathcal{N}$ be the corresponding completely positive and trace non-increasing map whenever the classical outputs have the coincidences $x_A=x_B$ and $z_A=z_B$. Ideally, the probability of this occurring is $\pi^{(XZ)}\equiv \Tr(\mathcal{N}(\overline{\rho})) < p^{(XZ)}$. Upon obtaining an estimate $\hat{\rho}$, we can evaluate the empirical counterpart $\hat{\pi}^{(XZ)}\equiv \Tr(\mathcal{N}(\hat{\rho}))\leq \hat{p}^{(XZ)}$. Propagating from \eqref{eqn:conc_bound_trace_dist_final}, we find that
\begin{equation}
    \Pr(D\!\left(\frac{\mathcal{N}(\hat{\rho})}{\hat{\pi}^{(XZ)}},\frac{\mathcal{N}(\overline{\rho})}{\pi^{(XZ)}}\right)\geq \varepsilon_{d}) \leq \delta_T,
\end{equation}
where $\abs{\smash[t]{\hat{\pi}^{(XZ)} - \pi^{(XZ)}}}\leq 2\varepsilon_T$, and $\varepsilon_{d} \equiv 2\varepsilon_T/(\hat{\pi}^{(XZ)} - 2\varepsilon_T)$ is now the additive error corresponding to the diagonal part ${\mathcal{N}(\overline{\rho})}/{\pi^{(XZ)}}$ of the distilled state.

\begin{figure*}[tp]
    \centering
    \subfloat[Tomography vs. Double selection Disti-Mator]{\includegraphics[width=0.46\textwidth]{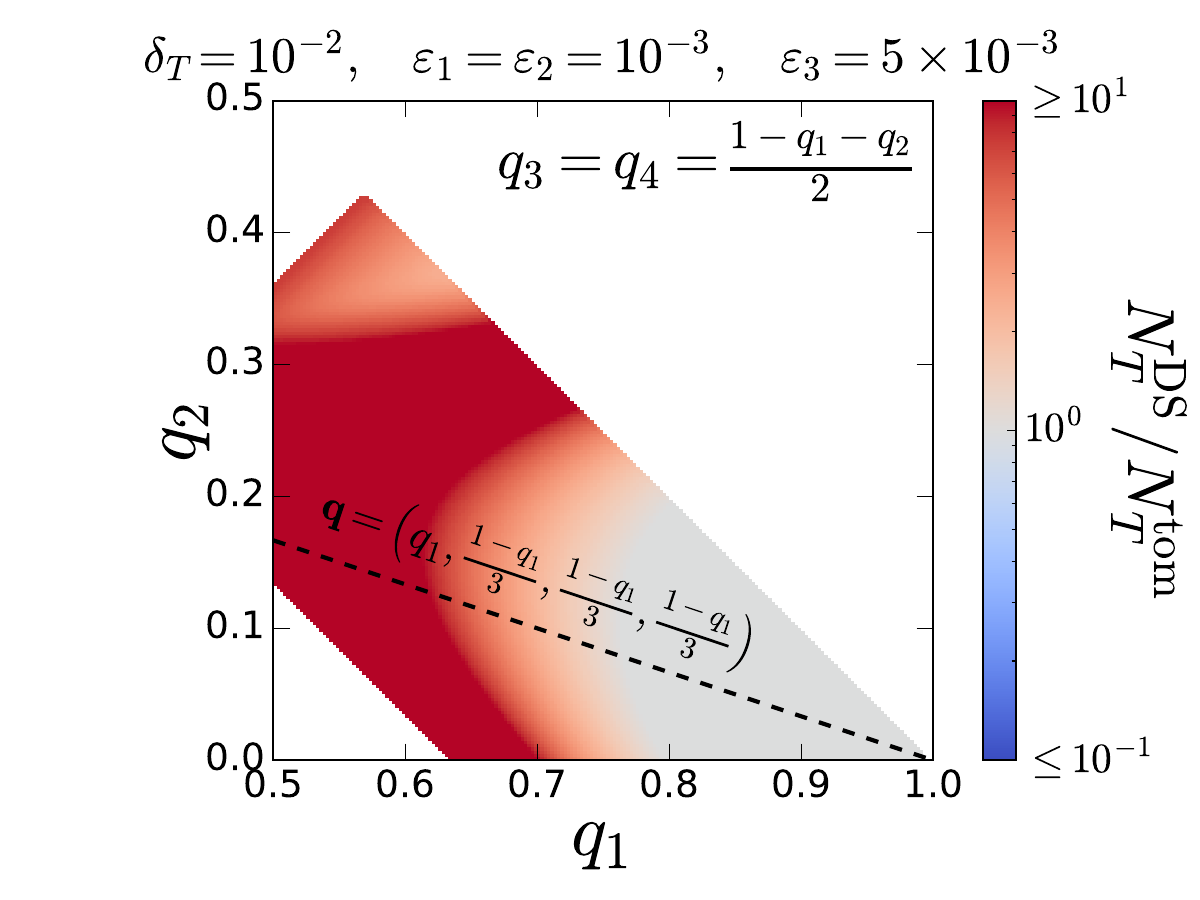}%
    \label{fig:complexity-tomography-double-selection-hoeffding}}
    \hfil
    \subfloat[Original Disti-Mator \cite{casapao2024distimator} vs. Double selection Disti-Mator]{\includegraphics[width=0.46\textwidth]{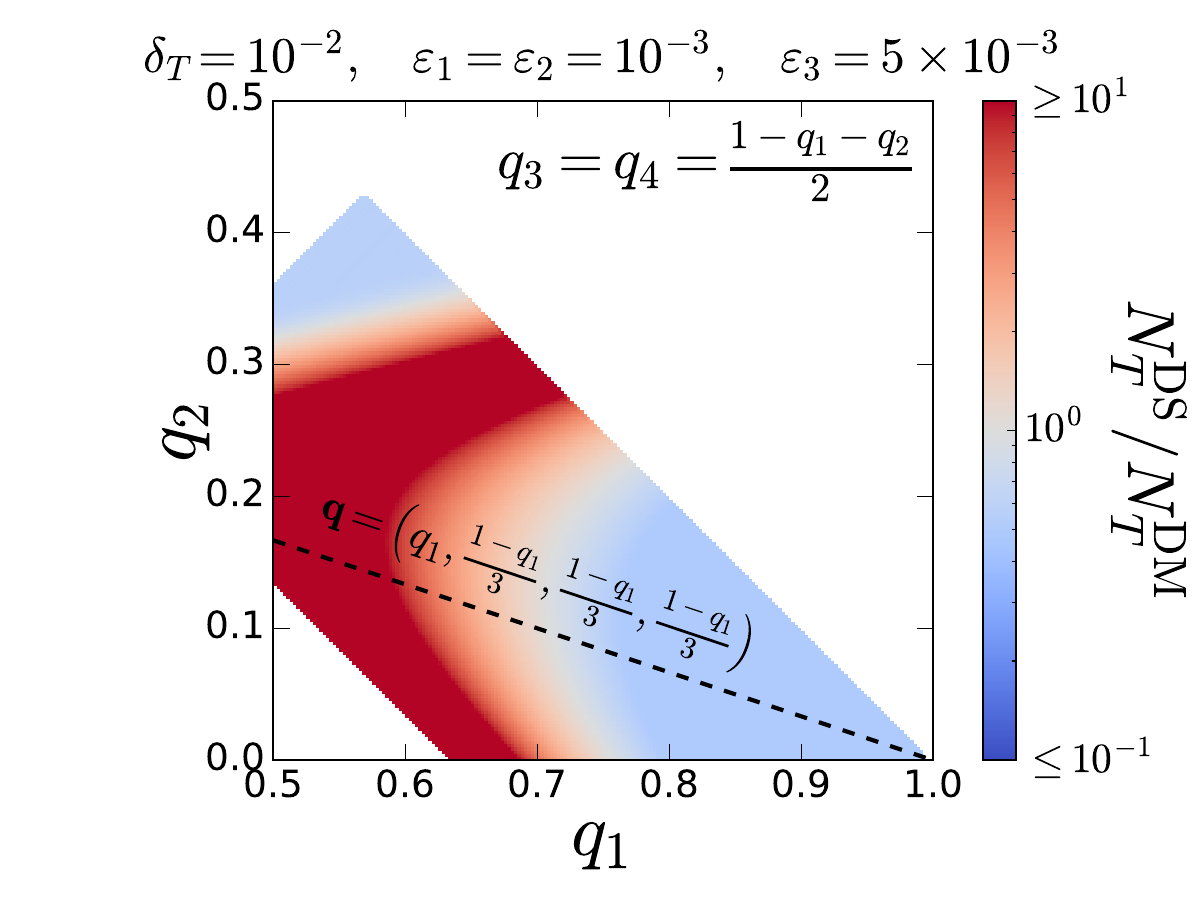}%
    \label{fig:complexity-double-selection-hoeffding}}
    \caption{
    (a)
    Ratio of the total number of noisy Bell pairs $N_{T}^{\mathrm{DS}}$ pessimistically needed for the double selection-based Disti-Mator, and the total number $N_{T}^{\mathrm{tom}}$ for a tomographic protocol described in the text.
    (b) 
    Ratio of the total number of noisy Bell pairs $N_{T}^{\mathrm{DS}}$ pessimistically needed for the double selection-based Disti-Mator, and the total $N_{T}^{\mathrm{DM}}$ for the original Disti-Mator in \cite{casapao2024distimator}, assuming that each distillation protocol in the original Disti-Mator uses the same total number of pairs.
    We fix the failure probability of the trace distance being greater than $\varepsilon_T$ to be $\delta_T = 10^{-2}$. Each method promises an estimate $\hat{x}_i$ with additive errors $\varepsilon_1=\varepsilon_2= 10^{-3}$, and $\varepsilon_3=5\times 10^{-3}$. We define the axes as the Bell-diagonal parameters $q_1$ (with no Pauli error) and $q_2$ (with Pauli error $Z$ only), and we investigate the collection of noisy Bell pairs that satisfy $q_3 = q_4 = (1-q_1-q_2)/2$ (with Pauli errors $X$ and $XZ$, respectively). The white dashed line shows the set of Werner states, in which the Pauli errors are equally likely. Brighter colors indicate that our proposed double selection Disti-Mator offers a same order of magnitude in or a reduction in resource cost when compared to both tomography and the original Disti-Mator, while darker colors indicate that the alternative approach is more preferrable.}
    \label{fig:sample_complexity}
\end{figure*}

\subsection{Numerical simulations}\label{subsec:numerical_simulations}

We now assess the performance of the double selection Disti-Mator through numerical simulations. We first compare the sample complexity of our protocol with a simple tomographic reconstruction that we can implement during state distribution. We describe this protocol as follows. Here, this (single-copy) tomographic protocol is implemented with local measurements only. Notice that we can write any $\overline{\rho}(\mathbf{q})$ as $\overline{\rho}({\mathbf{q}}) = (I\otimes I + \sum_{\sigma=X,Y,Z} t_{\sigma}({\mathbf{q}})\sigma\otimes\sigma)/4$, where $t_{\sigma}({\mathbf{q}})$ is a distinct combination of convex coefficients. Hence, we only need to perform joint measurements in $X\otimes X$, $Y\otimes Y$, and $Z\otimes Z$, then classically communicate the results, and finally count all the coincidences in each. We further assume that we perform each joint measurement with the same number of noisy states $N^{\mathrm{tom}}$. We show the corresponding concentration bound for the trace distance following this tomographic protocol in Appendix~\ref{app:tomography}.

Figure~\ref{fig:complexity-tomography-double-selection-hoeffding} shows the ratio between the \textit{total} number of noisy Bell pairs pessimistically needed for a Disti-Mator based on double selection and the \textit{total} number needed for the tomographic protocol described above. Here, we take $\varepsilon_1=\varepsilon_2= 10^{-3}$, and $\varepsilon_3=5\times 10^{-3}$, and we fix the failure probability of the trace distance being greater than $\varepsilon_T$ to be $\delta_T = 10^{-2}$. We then evaluate $N$ using a root finding method on the equation $\delta_T=\sum_{i}\delta_i$. Moreover, we investigate the collection of noisy Bell pairs that satisfy $q_3 = q_4 = (1-q_1-q_2)/2$. There is nothing special about this choice of degrees of freedom, except that we only need to impose the condition $q_1>1/2$. However, this choice highlights the $N$-ratio in the Werner state regime (shown as a white dashed line), in which the Pauli errors are equally likely. 

We see that our estimator promises a comparable consumption of resources with the above tomographic protocol for noisy input states that already have a high fidelity $q_1$. In Figure~\ref{fig:complexity-tomography-double-selection-hoeffding}, this is shown in gray, which corresponds to a ratio close to one. We also observe an obvious disadvantage for states whose fidelity is far from one, shown in red, which corresponds to a ratio larger than one. Unlike in tomography, in which all samples are fully consumed for estimation, there is some probability of regaining distilled states if the network users instead opt for estimating states using the statistics from double selection itself.

We also compare the complexity of the double selection-based Disti-Mator with the original Disti-Mator in \cite{casapao2024distimator}. Figure~\ref{fig:complexity-double-selection-hoeffding} shows the ratio between the total number of noisy Bell pairs needed for a Disti-Mator based on double selection, and the total number needed for the original Disti-Mator in \cite{casapao2024distimator}, assuming that each distillation protocol in the original estimator uses the same total number of states (which is an optimal choice for estimating Werner states). We evaluated the ratio by a procedure similar to that described earlier. We see that using our new estimator consumes fewer input states to estimate a considerable region of diagonal elements $\mathbf{q}$ considered here than the original Disti-Mator and can therefore be more resource efficient (the blue region, as shown in Figure~\ref{fig:complexity-double-selection-hoeffding}). It is worth noting that for double selection, there is a chance of regaining one distilled state after consuming three noisy pairs, in contrast with the chance of regaining at most three distilled pairs (with different resulting Bell parameters) from the three distillation protocols used in the original Disti-Mator. 

\begin{figure*}[t]
    \centering
    \subfloat[$\mathbf{q} = (0.62,0.15,0.05,0.18)$]{\includegraphics[width=0.43\textwidth]{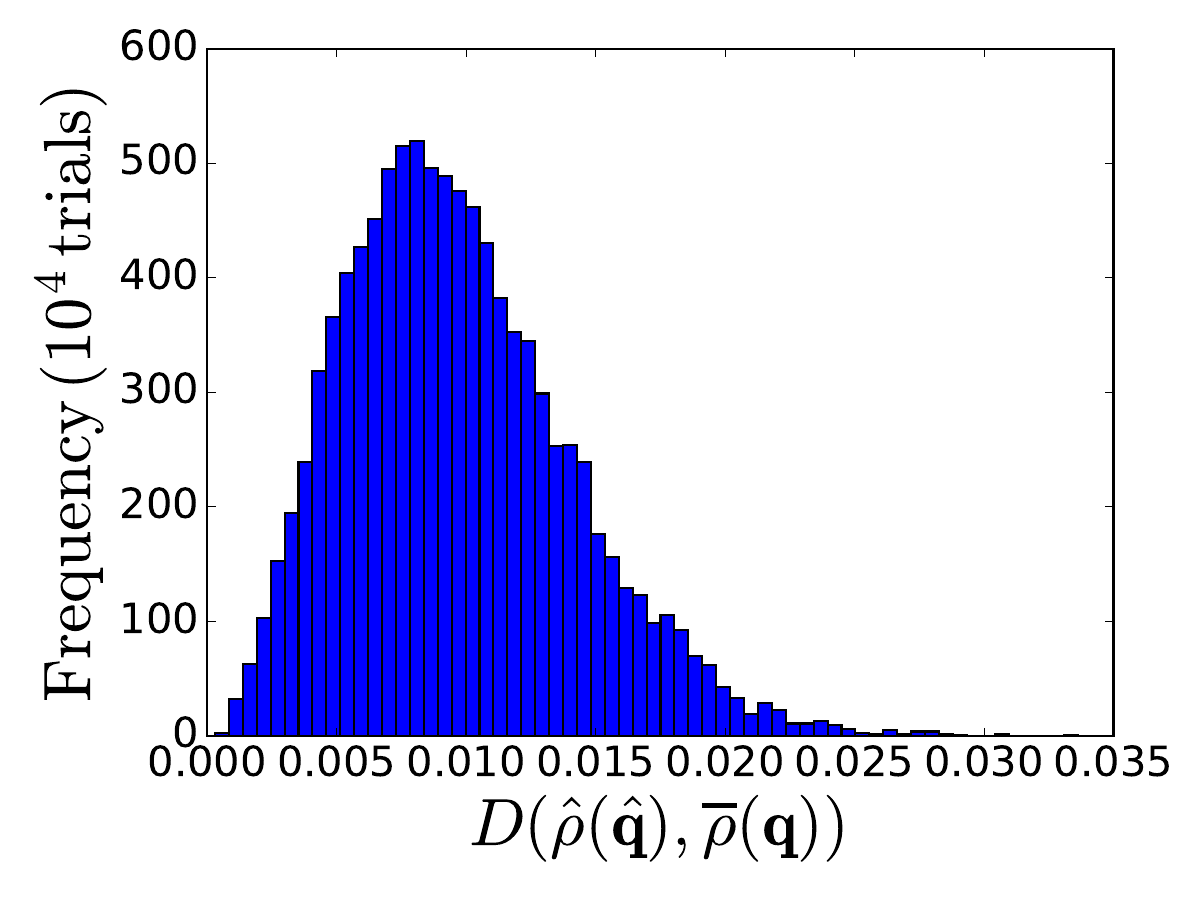}%
    \label{fig:numerical-simulation-histogram-lossy}}
    \hfil
    \subfloat[$\mathbf{q} = (0.88,0.02,0.05,0.05)$]{\includegraphics[width=0.43\textwidth]{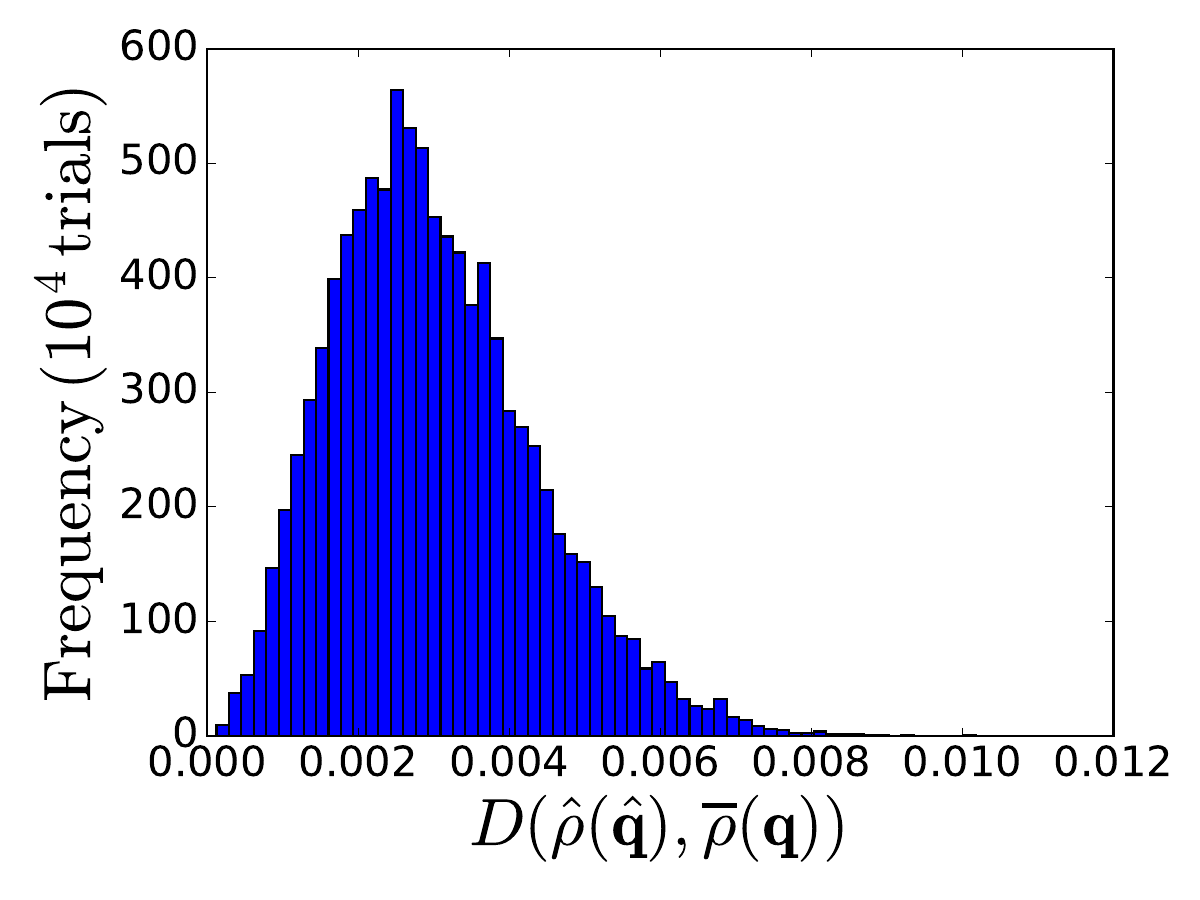}%
    \label{fig:numerical-simulation-histogram-high}}
    \caption{
    Histogram of the trace distances $D(\hat{\rho}(\hat{\mathbf{q}}),\overline{\rho}(\mathbf{q}))$, where each trial consumes $N=10^4$ triplets of noisy Bell pairs. Each noisy Bell pair is described by the vector $\mathbf{q}$ given above. The width of the bins are calculated based on the Freedman-Diaconis rule. The trace distance has a mean $\overline{D}$ and standard deviation $\sigma_D$ of: 
    (a)
    $\overline{D}\approx 0.010$ and $\sigma_D\approx 0.004$;
    (b) 
    $\overline{D}\approx 0.003$ and $\sigma_D\approx 0.001$.
    The positively skewed histograms show that our estimation method can accurately predict the initial state given only some typical number of measurements $N$ during implementation. Observing case (b), we expect that our method can predict high-fidelity input states extremely well.
    }
    \label{fig:histogram}
\end{figure*}

\subsection{Feasibility tests}\label{subsec:feasibility}

We now assess the feasibility of our proposed estimation scheme during a potential implementation. For our investigation, we first consider the typical values found in \cite{sakuma2024optical} to produce end-to-end fidelities with the currently available optical interconnect architecture. We describe the input noisy Bell pairs with parameters $\mathbf{q} = (0.62, 0.15, 0.05, 0.18)$, where $q_1= 0.62$ is the end-to-end fidelity, and the Bell pairs can be generated stably at an average rate of $5$ Hz. Whenever in this particular setting a double selection protocol is needed after establishing end-to-end entanglement, our proposed estimator can act as a real-time monitoring tool of the distributed Bell pairs. In an ideal scenario, we want our estimator to successfully estimate the pairs in a short amount of time. Given the above generation rate, we can reasonably set a time interval for the estimation to run on the order of minutes. Hence, as a feasibility test, we consider executing our estimation scheme on $N = 10^4$ triplets of Bell pairs. Taking into account the near-future improvements in the generation rate, we expect our scheme to obtain an estimate much faster with the given $N$ so long as the double selection protocol can be implemented efficiently.

Figure~\ref{fig:numerical-simulation-histogram-lossy} shows a histogram of the resulting trace distances with the given $\mathbf{q}$ above. The trace distance has a mean of about $0.010$ and a standard deviation of about $0.004$. The histogram reveals that the estimator performs well even with the relatively small $N$ given above. We observe that the distribution is unimodal and positively skewed, and the trace distances are typically no greater than $0.022$ (which is about three standard deviations from the mean). 

We also consider executing our estimation scheme with the same number of $N$ but with high-fidelity input states instead. States with high fidelities are already achievable with near-term devices, such as those demonstrated in \cite{Krutyanskiy2023entanglement,Knaut2024entanglement}, but with much smaller generation rates. Figure~\ref{fig:numerical-simulation-histogram-high} shows a histogram of the resulting trace distances given the Bell parameters $\mathbf{q} = (0.88,0.02,0.05,0.05)$. We still observe a well-behaved distribution of trace distances. However, an interesting difference from the histogram in Figure~\ref{fig:numerical-simulation-histogram-lossy} is that the trace distances are more tightly clustered around the mean $0.003$ with a smaller standard deviation. We also see that this mean is an order of magnitude smaller than the previous figure. That is, the earlier $\mathbf{q}$ case is more susceptible to shot noise. Moreover, we see that it is unlikely that the trace distance goes beyond $0.010$ (about seven standard deviations from the mean). Thus, we can be almost certain that we can predict $\mathbf{q}$ of this high-fidelity pair extremely well using our proposed estimator even with a relatively small $N$. We expect a similar behavior to apply to other input noisy Bell pairs with high enough fidelity.

%--------------------%

\section{Conclusions}\label{sec:conclusions}

In this work, we propose an estimator based on the double selection protocol. We demonstrate that the measurement statistics from this distillation protocol are sufficient to fully estimate the Bell-diagonal parameters of the undistilled states. We further show that the distilled states can also be estimated with additional post-processing. The performance of our estimator in terms of sample complexity is comparable to tomography with single-copy local measurements and can be better than the previously proposed distillation-based estimator in \cite{casapao2024distimator} if we wish to estimate high-fidelity input states. Furthermore, numerical simulations show that our estimator performs well under typical experimental circumstances.

An immediate open question is whether this estimation protocol is robust against noise induced by imperfections in the protocol, a topic that we leave for future investigation. We can also think of this work as part of the overarching problem of designing a parameter estimation strategy from an arbitrary distillation protocol, preferably without using any more resources beyond what is required for the protocol. We saw that the measurement statistics from double selection have relatively simple expressions which we have exploited to easily extract the Bell-diagonal elements. For more complex distillation protocols such as in \cite{krastanov2019optimized,goodenough2024near,Miguel-Ramiro2024improving}, designing an estimation recipe remains an avenue for further research. 

%--------------------%

\appendices

\section{Concentration bounds for $x_1$ and $x_2$}\label{app:concbound_x1x2}

In this appendix, we provide the concentration bounds on the probability of estimating $x_1$ and $x_2$ with accuracies $\varepsilon_1$ and $\varepsilon_2$, respectively. Let us first consider the concentration bound for $x_1$. For this case, we start with the expression for $x_1(p^{(Z)})$ in \eqref{eqn:doub_x1_noiseless}. Let $N$ be the size of the measurement data, giving an empirical probability $\hat{p}^{(Z)}$ that is at most $\varepsilon^{(Z)}$ away from the true coincidence probability $p^{(Z)}$. By Hoeffding's inequality \cite{hoeffding1963prob}, we have
\begin{multline}\label{eqn:pZ_hoeffding_bound}
    \Pr(\abs{\hat{p}^{(Z)} - p^{(Z)}}\geq \varepsilon^{(Z)})\\
    \leq 2\exp(-2N(\varepsilon^{(Z)})^2) \equiv \delta^{(Z)},
\end{multline}
where we take $p^{(Z)} = \mathbb{E}[\hat{p}^{(Z)}]$. With this guarantee on $\hat{p}^{(Z)}$, we can say that with probability no less than $1-\delta^{(Z)}$, we have the upper bound on $x_1$:
\begin{multline}
    \mathcal{U}_{x_1} \equiv \frac{1}{2}\left(1+ \left(2\min\left\{\hat{p}^{(Z)}+\varepsilon^{(Z)},1\right\}-1\right)^{1/3}\right)\\
    \geq x_1(p^{(Z)}).
\end{multline}
On the other hand, we have the lower bound
\begin{multline}
    \mathcal{L}_{x_1} \equiv \frac{1}{2}\left(1+ \left(2\max\left\{\hat{p}^{(Z)}-\varepsilon^{(Z)},\frac{1}{2}\right\}-1\right)^{1/3}\right)\\
    \leq x_1(p^{(Z)}).
\end{multline}
Then, we must satisfy
\begin{equation}
    x_1 - \hat{x}_1 \leq \mathcal{U}_{x_1} - x_1(\hat{p}^{(Z)}) \equiv \varepsilon_{1,R}, 
\end{equation}
and
\begin{equation}
    \hat{x}_1 - {x}_1 \leq x_1(\hat{p}^{(Z)}) - \mathcal{L}_{x_1} \equiv \varepsilon_{1,L},
\end{equation}
where we use the assumption $\hat{x}_1 \equiv x_1(\hat{p}^{(Z)})$ of the estimator. 

Thus, we have the probability inequality
\begin{multline}
    \Pr(\abs{\hat{p}^{(Z)} - p^{(Z)}}\leq \varepsilon^{(Z)}) \\
    \leq \Pr(-\varepsilon_{1,L}\leq x_1 - \hat{x}_1 \leq \varepsilon_{1,R}).
\end{multline}
We can simplify the expression by bounding the right-hand side of the above inequality with $\Pr(\abs{x_1 - \hat{x}_1}\leq \varepsilon_1)$, where $\varepsilon_1 \equiv \max\{\varepsilon_{1,L},\varepsilon_{1,R}\}$. Using the Hoeffding bound in \eqref{eqn:pZ_hoeffding_bound}, we have the guarantee
\begin{equation}
    \Pr(\abs{x_1 - \hat{x}_1}\leq \varepsilon_1) \geq 1 - \delta^{(Z)} \equiv 1 - \delta_1.
\end{equation}
Thus, we obtain a guarantee on the probability of estimating $x_1$ with accuracy $\varepsilon_1$. 

Next, we consider $x_2$. For this, we start with the expression for $x_2(p^{(X)})$ in \eqref{eqn:doub_x2_noiseless}. Given $N$ measurement results, we let $\hat{p}^{(X)}$ be the empirical coincidence probability that is at most $\varepsilon^{(X)}$ away from $p^{(X)}$. Hoeffding's inequality guarantees that
\begin{multline}\label{eqn:pX_hoeffding_bound}
    \Pr(\abs{\hat{p}^{(X)} - p^{(X)}}\geq \varepsilon^{(X)})\\
    \leq 2\exp(-2N(\varepsilon^{(X)})^2) \equiv \delta^{(X)},
\end{multline}
where $p^{(X)} = \mathbb{E}[\hat{p}^{(X)}]$. We proceed in a similar way to the case for $p^{(Z)}$. We say that with probability no less than $1-\delta^{(X)}$, we have the upper bound on $x_2$ given as
\begin{multline}
    \mathcal{U}_{x_2} \equiv \frac{1}{2}\left(1+ \left(2\min\left\{\hat{p}^{(X)}+\varepsilon^{(X)},1\right\}-1\right)^{1/2}\right)\\
    \geq x_2(p^{(X)}),
\end{multline}
and the lower bound given as
\begin{multline}
    \mathcal{L}_{x_2} \equiv \frac{1}{2}\left(1+ \left(2\max\left\{\hat{p}^{(X)}-\varepsilon^{(X)},\frac{1}{2}\right\}-1\right)^{1/2}\right)\\
    \leq x_2(p^{(X)}).
\end{multline}
Then, we must satisfy
\begin{equation}
    x_2 - \hat{x}_2 \leq \mathcal{U}_{x_2} - x_2(\hat{p}^{(X)}) \equiv \varepsilon_{2,R},
\end{equation}
and
\begin{equation}
    \hat{x}_2 - {x}_2 \leq x_2(\hat{p}^{(X)}) - \mathcal{L}_{x_2} \equiv \varepsilon_{2,L},
\end{equation}
where we use the assumption $\hat{x}_2 \equiv x_2(\hat{p}^{(X)})$ of the estimator. 

Thus, we have the probability inequality
\begin{multline}
    \Pr(\abs{\hat{p}^{(X)} - p^{(X)}}\leq \varepsilon^{(X)}) \\
    \leq \Pr(-\varepsilon_{2,L}\leq x_2 - \hat{x}_2 \leq \varepsilon_{2,R}).
\end{multline}
Similarly to before, we can bound the right-hand side of this inequality with $\Pr(\abs{x_2 - \hat{x}_2}\leq \varepsilon_2)$, where $\varepsilon_2 \equiv \max\{\varepsilon_{2,L},\varepsilon_{2,R}\}$. Using the Hoeffding bound in \eqref{eqn:pX_hoeffding_bound}, we have 
\begin{equation}
    \Pr(\abs{x_2 - \hat{x}_2}\leq \varepsilon_2) \geq 1 - \delta^{(X)} \equiv 1 - \delta_2,
\end{equation}
giving us a guarantee on the probability of estimating $x_2$ with accuracy $\varepsilon_2$. 

\section{Concentration bound for $x_3$}\label{app:concbound_x3}

In this appendix, we provide a concentration bound on the probability of estimating $x_3$ with accuracy $\varepsilon_3$. To prove this bound, we begin with the expression for $x_3$ in \eqref{eqn:doub_x3_noiseless}. Unlike in Appendix~\ref{app:concbound_x1x2}, we treat the case of $x_3$ differently since it is a multivariate function of the coincidence probabilities. Our approach is to pessimistically bound $x_3({p}^{(Z)},{p}^{(X)},p^{(XZ)})$ using two single-variable functions $\mathcal{U}(p^{(XZ)})$ and $\mathcal{L}(p^{(XZ)})$ assuming that $\hat{p}^{(Z)}$ and $\hat{p}^{(X)}$ successfully estimate the true coincidence probabilities ${p}^{(Z)}$ and ${p}^{(X)}$ with high probability. The errors for these empirical quantities are defined in Appendix~\ref{app:concbound_x1x2}.

Let $N$ be the size of the measurement data, giving an empirical probability $\hat{p}^{(XZ)}$ that is at most $\varepsilon^{(XZ)}$ away from the true coincidence probability $p^{(XZ)}$. By Hoeffding's inequality, we have
\begin{multline}\label{eqn:pXZ_hoeffding_bound}
    \Pr(\abs{\hat{p}^{(XZ)} - p^{(XZ)}}\geq \varepsilon^{(XZ)})\\
    \leq 2\exp(-2N(\varepsilon^{(XZ)})^2) \equiv \delta^{(XZ)},
\end{multline} 
where $p^{(XZ)} = \mathbb{E}[\hat{p}^{(XZ)}]$. We then have the upper bound to $x_3$:
\begin{multline}
    \mathcal{U}_{x_3}(p^{(XZ)})\equiv\\
     x_3\!\left(\!\max\!\left\{\hat{p}^{(Z)}-\varepsilon^{(Z)},\frac{1}{2}\right\},\max\!\left\{\hat{p}^{(X)}-\varepsilon^{(X)},\frac{1}{2}\right\},p^{(XZ)}\right)\\
    \geq x_3({p}^{(Z)},{p}^{(X)},p^{(XZ)}).
\end{multline}
On the other hand, we have the lower bound
\begin{multline}
    \mathcal{L}_{x_3}(p^{(XZ)})\equiv\\
     x_3\!\left(\min\!\left\{\hat{p}^{(Z)}+\varepsilon^{(Z)},1\right\},\min\!\left\{\hat{p}^{(X)}+\varepsilon^{(X)},1\right\},p^{(XZ)}\right)\\
    \leq x_3({p}^{(Z)},{p}^{(X)},p^{(XZ)}).
\end{multline}
We further assume that $\hat{p}^{(XZ)}$ lies within the union of the domains of $\mathcal{U}_{x_3}$ and $\mathcal{L}_{x_3}$ over the real numbers. Then, we must satisfy
\begin{multline}
    x_3 - \hat{x}_3 \\
    \leq \mathcal{U}_{x_3}\!\left(\min\left\{\hat{p}^{(XZ)}+\varepsilon^{(XZ)},1\right\}\right) - x_3(\hat{p}^{(Z)},\hat{p}^{(X)},\hat{p}^{(XZ)})\\
    \equiv \varepsilon_{3,R},
\end{multline}
and
\begin{multline}
    \hat{x}_3 - {x}_3 \\
    \leq x_3(\hat{p}^{(Z)},\hat{p}^{(X)},\hat{p}^{(XZ)}) - \mathcal{L}_{x_3}\!\left(\max\left\{\hat{p}^{(XZ)}-\varepsilon^{(XZ)},\frac{1}{4}\right\}\right)\\
    \equiv \varepsilon_{3,L},
\end{multline}
where we use the assumption $\hat{x}_3 \equiv x_3(\hat{p}^{(Z)},\hat{p}^{(X)},\hat{p}^{(XZ)})$ of the estimator. We also note that $p^{(XZ)} = 1/4$ occurs when $x_i=1/2$ for all $i$, which corresponds to the maximally mixed state.

Thus, we arrive at the following probability inequality:
\begin{multline}
    \Pr(\bigwedge_{j=X,Z,XZ}\left\{\abs{\hat{p}^{(j)} - p^{(j)}}\leq \varepsilon^{(j)}\right\})\\
    \leq \Pr(-\varepsilon_{3,L}\leq x_3 - \hat{x}_3 \leq \varepsilon_{3,R}).
\end{multline}
We can bound the right-hand side of this inequality with $\Pr(\abs{x_3 - \hat{x}_3}\leq \varepsilon_3)$, where we denote the additive error $\varepsilon_3 \equiv \max\{\varepsilon_{3,L},\varepsilon_{3,R}\}$. We can also bound the probability of the above logical conjunction using the intersection bound on probabilities (via the union bound). We therefore obtain the concentration bound
\begin{equation}
    \Pr(\abs{\hat{x}_3-x_3}\leq \varepsilon_3)
    \geq 1- \sum_{j=X,Z,XZ}\delta^{(j)} \equiv 1-\delta_3,
\end{equation}
where we used \eqref{eqn:pZ_hoeffding_bound}, \eqref{eqn:pX_hoeffding_bound}, and \eqref{eqn:pXZ_hoeffding_bound}. Hence, we obtain a guarantee on the probability of estimating $x_3$ with accuracy $\varepsilon_3$.

\section{Concentration bound for the trace distance}\label{app:conc_trace_distance}

We finally present the concentration bound for the trace distance $D(\hat{\rho}(\hat{\mathbf{q}}),\overline{\rho}(\mathbf{q}))$ between the true Bell-diagonal $\overline{\rho}(\mathbf{q})$ and our estimation $\hat{\rho}(\hat{\mathbf{q}})$. We recall that the trace distance is defined as
\begin{equation}
    D(\hat{\rho}(\hat{\mathbf{q}}),\overline{\rho}(\mathbf{q})) \equiv \frac{1}{2}\Vert \hat{\rho}(\hat{\mathbf{q}}) - \overline{\rho}(\mathbf{q}) \Vert_1,
\end{equation}
where $\Vert \cdot\Vert_1$ is the matrix trace norm. Since both $\hat{\rho}$ and $\overline{\rho}$ are in Bell-diagonal form, we can simplify the above equation into
\begin{equation}
    D(\hat{\rho}(\hat{\mathbf{q}}),\overline{\rho}(\mathbf{q})) = \frac{1}{2}\sum_{k=1}^4 \abs{\hat{q}_k - q_k} = \frac{1}{2}\Vert \hat{\mathbf{q}} - \mathbf{q} \Vert_1,
\end{equation}
where $\Vert \cdot\Vert_1$ now denotes the usual $\ell_1$-norm.

Assume that we have achieved conditions $\abs{\hat{x}_i - x_i}\leq \varepsilon_i$ for all $i$, where each $\varepsilon_i$ is determined by $\varepsilon^{(Z)}$, $\varepsilon^{(X)}$, and $\varepsilon^{(XZ)}$ following Appendices~\ref{app:concbound_x1x2} and~\ref{app:concbound_x3}. With the above conditions, as well as transformation rule in \eqref{eqn:q-x_transformation} and the triangle inequality, we have
\begin{multline}
    D(\hat{\rho}(\hat{\mathbf{q}}),\overline{\rho}(\mathbf{q})) \leq \frac{1}{2}\sum_{k=1}^4 \frac{1}{2}\left(\sum_{i=1}^3\abs{\hat{x}_i-x_i}\right) \\
    = \sum_{i=1}^3 \abs{\hat{x}_i-x_i} \leq \sum_{i=1}^3 \varepsilon_i = \varepsilon_T.
\end{multline}
Thus, we obtain the probability inequality
\begin{equation}
    \Pr(\bigwedge_{i=1}^3\{\abs{\hat{x}_i - x_i}\leq \varepsilon_i\}) \leq \Pr(D(\hat{\rho}(\hat{\mathbf{q}}),\overline{\rho}(\mathbf{q}))\leq \varepsilon_T).
\end{equation}
We can bound the probability of this logical conjunction by using the intersection bound on probabilities. With the concentration bounds $\delta_i$'s obtained in Appendices~\ref{app:concbound_x1x2} and~\ref{app:concbound_x3}, we ultimately have
\begin{equation}\label{eqn:app_trace_dist_bound}
    \Pr(D(\hat{\rho}(\hat{\mathbf{q}}),\overline{\rho}(\mathbf{q}))\geq \varepsilon_T)\leq  \sum_{i=1}^3 \delta_i= \delta_T.
\end{equation}

\section{Sample complexity of the proposed estimator}\label{app:complexity}

We provide a short discussion of the sample complexity of our proposed estimator. We follow the same notation as in the previous Appendices. As stated in the main text, we assume that all $\varepsilon^{(j)} = {\varepsilon}$ for $j=X,Z,XZ$, and therefore $\delta^{(j)}=\delta$ for all $j$. We also describe all concentration bounds with the same measurement data on $N$ triplets used during the double selection protocol. Thus,
\begin{equation}
    \delta_T = 2\delta^{(Z)} + 2\delta^{(X)} + \delta^{(XZ)}
    =5\delta= 10\exp(-2N{\varepsilon}^2)
\end{equation}
so that we have we the upper bound 
\begin{equation}
    N = \mathcal{O}\left(\frac{\log(1/\delta)}{{\varepsilon}^2}\right)
\end{equation}
in terms of the Hoeffding parameters $\varepsilon,\delta$. 

We can also rewrite this bound in terms of the parameters $\varepsilon_T,\delta_T$ in \eqref{eqn:app_trace_dist_bound}. We see from the above calculations that $\delta_T$ is of the same order as $\delta$. Now, we observe after an application of the binomial series that $\varepsilon_i= \mathcal{O}(\varepsilon)$ for all $i$ as $\varepsilon\rightarrow0^+$, with the $\varepsilon_i$'s defined in Appendices~\ref{app:concbound_x1x2} and~\ref{app:concbound_x3}, and the calculation for $\varepsilon_3$ being slightly more involved than the other two. Therefore, we have
\begin{equation}
    N = \mathcal{O}\left(\frac{\log(1/\delta_T)}{\varepsilon_T^2}\right)
\end{equation}
since $\varepsilon_T = \mathcal{O}(\varepsilon)$.

\section{Concentration bound associated with a single-copy local measurement-based tomography}\label{app:tomography}

Here, we calculate the corresponding concentration bound for the trace distance using a local measurement-based tomographic protocol described in the main text. We recall that any $\overline{\rho}(\mathbf{q})$ can be written as $\overline{\rho}({\mathbf{q}}) = (I\otimes I + \sum_{\sigma=X,Y,Z} t_{\sigma}({\mathbf{q}})\sigma\otimes\sigma)/4$, where $t_{\sigma}({\mathbf{q}})$ is a distinct combination of convex coefficients. In particular, these coefficients are $t_X({\mathbf{q}})=q_1-q_2+q_3-q_4$, $t_Y({\mathbf{q}}) = -q_1+q_2+q_3-q_4$, and $t_Z({\mathbf{q}})= q_1+q_2-q_3-q_4$. 

Let $p^{(\sigma)}_{\mathrm{tom}}$ be the coincidence probability when Alice and Bob jointly measure $\sigma\otimes\sigma$, where $\sigma=X,Y,Z$. We find that
\begin{equation}
    p^{(Z)}_{\mathrm{tom}} = x_1,\quad p^{(X)}_{\mathrm{tom}} = x_2,\quad p^{(Y)}_{\mathrm{tom}} = 1-x_3.
\end{equation}
Hence, we guarantee conditions $\abs{\hat{x}_i - x_i}\leq \varepsilon_i$ for all $i$ if the coincidence probabilities $p^{(Z)}_{\mathrm{tom}}$, $p^{(X)}_{\mathrm{tom}}$, and $p^{(Y)}_{\mathrm{tom}}$ can be estimated with errors $\varepsilon_1$, $\varepsilon_2$, and $\varepsilon_3$, respectively. Following a straightforward application of Hoeffding's inequality (similar to that in Appendix~\ref{app:conc_trace_distance}), we obtain
\begin{multline}
    \Pr(D(\hat{\rho}(\hat{\mathbf{q}}),\overline{\rho}(\mathbf{q}))\geq \varepsilon_T)\\
    \leq \sum_{i=1}^3 2\exp(-2N_i^{\mathrm{tom}}\varepsilon_i^2)
    \equiv \delta_T,
\end{multline}
where $N_i^{\mathrm{tom}}$ is the resource cost needed for a given $i$.

\bibliographystyle{IEEEtran}
\bibliography{reference.bib}

\end{document}